\documentclass[aps,showpacs,PRApplied,twocolumn,superscriptaddress,apsrev,english]{revtex4-1}
\usepackage{multirow}
\usepackage{color}
\usepackage{amsfonts}
\usepackage{amsmath}
\usepackage{mathrsfs}
\usepackage{amssymb}
\usepackage{wasysym}
\usepackage{graphicx}
\usepackage{babel}
\usepackage{graphicx}
\usepackage{paralist}
\usepackage{indentfirst}
\usepackage{subfigure}
\usepackage{hyperref}
\usepackage{ulem}
\usepackage{soul}
\usepackage{cancel}
\usepackage{CJK}

\begin{document}
\begin{CJK*}{GBK}{}
\title{Lifetime Determination of the $5s5p$ ${}^{3}P^{\rm{o}}_{0}$ Metastable State in ${}^{87}$Sr from the Electric Dipole Matrix Element}
\author{Xiao-Tong Lu}
\thanks{These authors contributed equally to this work.}
\affiliation{National Time Service Center, Chinese Academy of Sciences, Xi'an 710600, China}
\author{Feng Guo}
\thanks{These authors contributed equally to this work.}
\affiliation{National Time Service Center, Chinese Academy of Sciences, Xi'an 710600, China}
\author{Yan-Yan Liu}
\affiliation{National Time Service Center, Chinese Academy of Sciences, Xi'an 710600, China}
\author{Jing-Jing Xia}
\affiliation{National Time Service Center, Chinese Academy of Sciences, Xi'an 710600, China}
\affiliation{School of Astronomy and Space Science, University of Chinese Academy of Sciences, Beijing 100049, China}
\author{Guo-Dong Zhao}
\affiliation{National Time Service Center, Chinese Academy of Sciences, Xi'an 710600, China}
\affiliation{School of Astronomy and Space Science, University of Chinese Academy of Sciences, Beijing 100049, China}
\author{Ying-Xin Chen}
\affiliation{National Time Service Center, Chinese Academy of Sciences, Xi'an 710600, China}
\affiliation{School of Astronomy and Space Science, University of Chinese Academy of Sciences, Beijing 100049, China}
\author{Ye-Bing Wang}
\affiliation{National Time Service Center, Chinese Academy of Sciences, Xi'an 710600, China}
\author{Ben-Quan Lu}
\email{lubenquan@cqu.edu.cn}
\affiliation{National Time Service Center, Chinese Academy of Sciences, Xi'an 710600, China}
\author{Hong Chang}
\email{changhong@ntsc.ac.cn}
\affiliation{National Time Service Center, Chinese Academy of Sciences, Xi'an 710600, China}
\affiliation{School of Astronomy and Space Science, University of Chinese Academy of Sciences, Beijing 100049, China}
\begin{abstract}
We report a measurement of the radiative lifetime of the $5s5p \; {}^{\rm{3}}P^{\rm{o}}_{\rm{0}}$ metastable state in ${}^{87}$Sr, which is coupled to the 5$s^{\rm{2}} \;$ ${}^{\rm{1}}S_{\rm{0}}$ ground state via a hyperfine-induced electric dipole transition. The radiative lifetime is determined to be 151.4(48) s, in good agreement with theoretical results. Our approach relies on accurate measurements of laser intensity and free-space Rabi frequency, enabling lifetime measurements of any excited state and particularly suitable for long-lived states.
\end{abstract}


\maketitle
\end{CJK*}

\section{Introduction}
Precisely determining the lifetime of electronic states serves as a crucial reference point for understanding atomic structure. As such, it plays a critical role in improving the performance of various quantum techniques \cite{Bothwell2022[1],Sanner2019[2],Takamoto2022[3],Drscher2021[4],Golovizin2019[5],Grotti2018[6],Lu2021[7],Daley2008[8]}. For example, the understanding of atomic structure can help to reduce the uncertainty of dynamical black-body-radiation shift, which limits the accuracy of the state-of-the-art ${}^{87}$Sr \cite{Bothwell2022[1]} and ${}^{171}$Yb \cite{McGrew2018[31]} optical lattice clocks and the corresponding correction relies on the ab-initio calculations \cite{Porsev2006,Safronova2013}. It is relatively straightforward to precisely measure the lifetime by directly observing the spontaneous decay \cite{Beloy2012[9],Olmschenk2009[10]}. However, when the lifetime of an excited state exceeds seconds, such as in the case of metastable states, the direct observation of spontaneous decay faces challenges due to competing processes. In such scenarios, careful consideration must be given to the evaluation of decaying rates arising from collisions and off-resonant laser radiation. Additionally, the long measurement period is necessary to ensure adequately small statistical uncertainty \cite{Yasuda2004[11],Dorscher2018[12]}.\ 

The natural lifetime of the metastable state could be precisely inferred from the cavity-enhanced dispersion measurements, where a transition with a well-known spontaneous emission rate provides a reference for the target level. This complicated method has been demonstrated for lifetime measurements of the $\left|5s5p \; {}^{3}P^{\rm{o}}_{0}\right\rangle$ state in ${}^{87}$Sr \cite{Muniz2021[13]}, achieving a measurement uncertainty of 2.5\%. However, it is worth noting that the result of 118(3) s obtained in their experiment is inconsistent with the prior measurement of 330(170) s \cite{Dorscher2018[12]}. This discrepancy highlights the necessity for additional investigation into the lifetime of the $\left|5s5p \; {}^{3}P^{\rm{o}}_{0}\right\rangle$ state in ${}^{87}$Sr. Alternatively, combining measurements of Rabi frequency and Stark shifts from the interrogation laser can extract the lifetime from the transition matrix element \cite{Hettrich2015[14]}. Recently, this method was used to measure the lifetime of the ${}^{2}F_{7/2}$ level in $\rm{{}^{171}Yb^{+}}$ with record precision for such a long lifetime in the $10^{7}$ s \cite{Lange2021[15]}. It appears that this technique reduces the challenging requirement for precise determination of the laser intensity. However, it is still necessary to determine the differential polarizability between states, and in certain cases, accurately measuring the laser intensity remains crucial \cite{Huntemann2016[23]}. \  

In this paper, we have developed a technique that enables accurate reconstruction of the two-dimensional laser intensity profile. This advancement allows for precise determination of the natural lifetime of metastable states through the measurement of the transition matrix element. By employing cold-atom ensemble as a sensor, we can effectively detect the distribution of laser intensity and accurately determine the central position of the laser beam profile by systematically adjusting its location. Combining this information with the result of the free-space Rabi frequency, the corresponding transition matrix element can be accurately determined. To demonstrate this technique, we investigate the hyperfine-induced E1 transition from the $ \left| 5s5p \; {}^{3}P^{\rm{o}}_{0},F^{\rm{e}}=I,M^{\rm{e}}_{\rm{F}}=+9/2 \right\rangle $  excited state to the $ \left| 5s^{2} \; {}^{1}S_{0},F^{\rm{g}}=I,M^{\rm{g}}_{\rm{F}}=+9/2 \right\rangle $  ground state in ${}^{87}$Sr (with a nuclear spin $I=9/2$). The natural lifetime of the excited $\left|5s5p \; {}^{3}P^{\rm{o}}_{0}\right\rangle$ state has been determined with 3.2\% uncertainty. \\

\section{Method}
The transition rate $ A (M^{\rm{g}}_{\rm{F}},M^{\rm{e}}_{\rm{F}})$ from an excited state $\left| \Gamma I J^{\rm{e}}F^{\rm{e}}M^{\rm{e}}_{\rm{F}}\right\rangle$ to a lower state (ground state) $\left| \Gamma I J^{\rm{g}}F^{\rm{g}}M^{\rm{g}}_{\rm{F}}\right\rangle$ can be written as \cite{Grumer2014[16]}

\begin{equation}
\begin{split}
&A (M^{\rm{g}}_{\rm{F}},M^{\rm{e}}_{\rm{F}}) \\
&= \frac{C^{\rm{\rho k}}}{\lambda^{\rm{2k+1}}} \sum_{\rm{q}} |\left\langle \Gamma I J^{\rm{g}}F^{\rm{g}}M^{\rm{g}}_{\rm{F}}| O^{\rm{\rho k}}_{\rm{q}} | \Gamma I J^{\rm{e}}F^{\rm{e}}M^{\rm{e}}_{\rm{F}} \right\rangle |^{2}. \label{eq1}
\end{split}
\end{equation}

$\left\langle \Gamma I J^{\rm{g}}F^{\rm{g}}M^{\rm{g}}_{\rm{F}}| O^{\rm{\rho k}}_{\rm{q}} | \Gamma I J^{\rm{e}}F^{\rm{e}}M^{\rm{e}}_{\rm{F}} \right\rangle $ is the transition matrix element between the ground state and the excited state. $O^{\rm{\rho k}}_{\rm{q}}$ indicates the multipole radiation-field tensor operator and is typically determined by the rank $k$, which corresponds to the angular momentum of the photon relative to atoms. $\rho=(-1)^{\rm{k}}$ or $(-1)^{\rm{k+1}}$ represents the parity for the electric (E$k$) or magnetic (M$k$) multipole transitions, respectively. $C^{\rm{\rho k}}$ is a constant depending on the transition type and $\lambda$ is the transition wavelength. In relation to the $\left| \Gamma I J^{\rm{e}}F^{\rm{e}}M^{\rm{e}}_{\rm{F}}\right\rangle$ $\rightarrow$ $\left| \Gamma I J^{\rm{g}}F^{\rm{g}}M^{\rm{g}}_{\rm{F}}\right\rangle$ transition, the free-space Rabi frequency, which indicates the strength of the interaction between atoms and the electromagnetic field used for interrogation, can be expressed by \cite{Scar1981[17]}

\begin{equation}
\Omega_{\rm{M^{eg}_{F}}}= |\left\langle \Gamma I J^{\rm{g}}F^{\rm{g}}M^{\rm{g}}_{\rm{F}}| O^{\rm{\rho k}} | \Gamma I J^{\rm{e}}F^{\rm{e}}M^{\rm{e}}_{\rm{F}} \right\rangle | \cdot \vec{E}/ \hbar. \label{eq2}
\end{equation}

$\hbar$ represents the reduced Planck constant, and $\vec{E}$ represents the electric field strength, which is directly related to the intensity ($I_{0}$) of the interrogation laser by $|\vec{E}|^2=2I_{0}n_{\rm{r}}/c_{0} n_{0} \varepsilon_{0}$. Here, $n_{\rm{r}}$ and $n_{0}$ denote the refractive index of the atom ensemble and the surrounding space, respectively. $c_{0}$ stands for the speed of light in a vacuum, while $\varepsilon_{0}$ represents the vacuum permittivity. Therefore, the transition rate $A (M^{\rm{g}}_{\rm{F}},M^{\rm{e}}_{\rm{F}})$ can be directly determined based on the measurements of $\Omega_{\rm{M^{eg}_{F}}}$ and $I_{0}$.  

\begin{figure}[tbp]\setlength{\abovecaptionskip}{-0.3mm}
\begin{center}
\includegraphics[width=85mm]{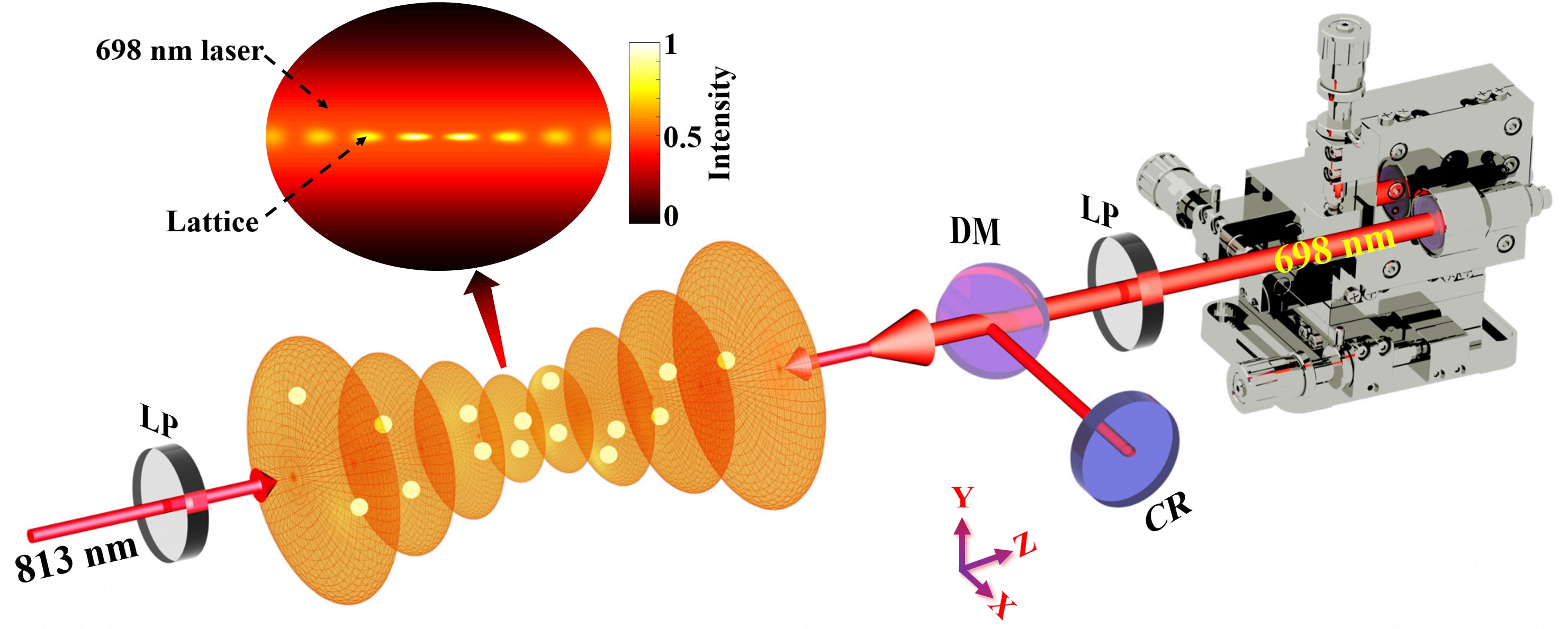}
\caption{\label{fig1} Experimental setup for optical lattice and clock transition detection. The lattice is created by overlapping the incident 813 nm laser beam with its retroreflected beam using a concave mirror (CR). The 698 nm interrogation laser beam is aligned with the lattice laser using a dichroic mirror (DM). Both lasers are polarized along the Y direction, and their polarizations are ensured using the linear polarizer (LP). The position of the 698 nm laser can be finely adjusted using a three-dimensional translation stage. The inset shows that the waist of the 698 nm laser is much larger than the lattice size, and atoms can only interact with a small portion of the 698 nm laser.}
\end{center}
\end{figure}\setlength{\belowcaptionskip}{-9.5mm}

\section{Experiment setup}
We present our method with the $\left| 5s5p \; {}^{3}P^{\rm{o}}_{0}, F^{\rm{e}}=9/2,M^{\rm{e}}_{\rm{F}}=+9/2\right\rangle$ $\rightarrow$ $\left| 5s^{2} \; {}^{1}S_{0}, F^{\rm{g}}=9/2, M^{\rm{g}}_{\rm{F}}=+9/2\right\rangle$ transition of ${}^{87}$Sr atoms. The experimental setup is based on our ${}^{87}$Sr transportable optical lattice clock \cite{Kong2020[18],Kong2021[19]}. Fig. \ref{fig1} shows the setup for the optical lattice and the clock transition detection. After undergoing two stages of laser cooling [see Appendix \ref{AP_a} for details], cold atoms are trapped within a horizontal one-dimensional optical lattice. The initial trap depth is set at 120 $E_{\rm{R}}$ ($E_{\rm{R}}$ is the recoil energy from a lattice photon). Subsequently, we employ an optical repumping technique to prepare the atoms in $\left|{}^{1}S_{0}, F^{\rm{g}}=I,M^{\rm{g}}_{\rm{F}}=+9/2\right\rangle$ ground state. In order to remove hotter atoms, we employ an energy-filtering method. This involves linearly reducing the lattice trap depth to 60 $E_{\rm{R}}$ over a period of 20 ms. We then wait for 10 ms before increasing the trap depth back up to 176 $E_{\rm{R}}$ within another 20 ms. Through this process, approximately 1000 atoms are retained within the lattice, while the hotter atoms are effectively removed. Following the filtering procedure, the axial temperature is measured to be 2.7 $\rm{\mu K}$ and the radial temperature is determined to be 3.9 $\rm{\mu K}$ using the sideband spectra \cite{Blatt2009[20]}.  

\begin{figure}[tbp]
	\begin{center}
		\includegraphics[width=70mm]{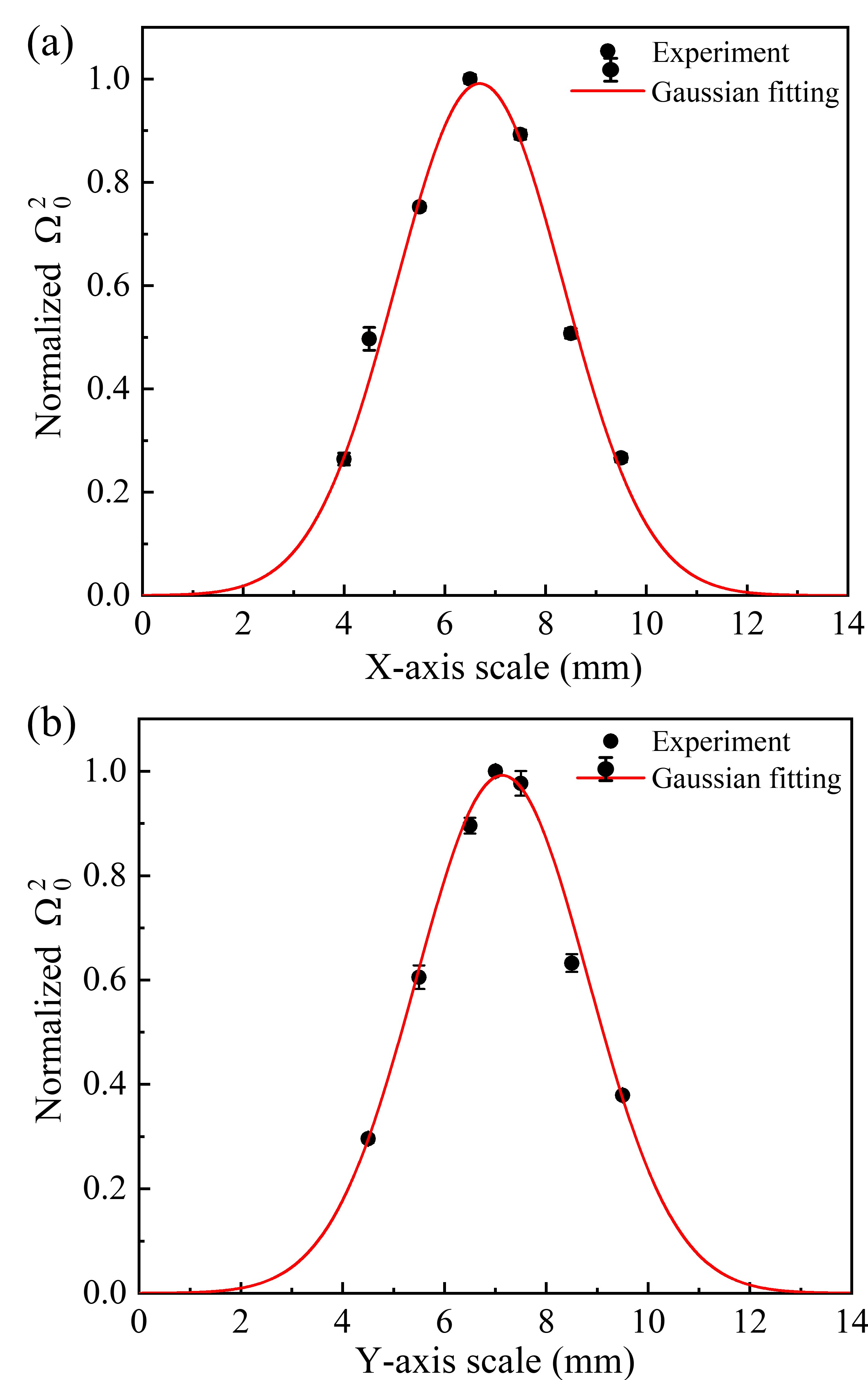}
		\caption{\label{fig2} The measurements of the square of the normalized free-space Rabi frequency $\Omega^{2}_{0}$ as a function of axis position. (a) $\Omega^{2}_{0}$ as a function of the X. (b) $\Omega^{2}_{0}$ as a function of the Y. The continuous lines show the Gaussian fittings. }
	\end{center}
\end{figure}

After preparing the quantum reference system, we proceed to interrogate the $\left|{}^{3}P^{\rm{o}}_{0},F^{\rm{e}}=9/2, M^{\rm{e}}_{\rm{F}}=+9/2\right\rangle$ $\rightarrow$ $\left|{}^{1}S_{0}, F^{\rm{g}}=9/2, M^{\rm{g}}_{\rm{F}}=+9/2\right\rangle$ transition using a 698 nm laser. The 698 nm laser is stabilized to an ultralow-expansion (ULE) cavity (with a fineness of 300,000) using the Pound-Drever-Hall (PDH) technique. To minimize frequency drift caused by changes in the cavity length, we implement a linear sweeping of the driving frequency of the acoustic optical modulator (AOM) and, effectively reduce the drift to below 1 mHz/s. The excitation fraction is determined using the normalized electronic shelving method \cite{Nagourney1986[21]}. After interrogating the 698 nm transition, the number of atoms in the $\left|{}^{1}S_{0}\right\rangle$ state ($N_{\rm{g}}$) is probed using a 461 nm laser pulse, while concurrently expelling these atoms from the lattice. Next, the atoms in the $\left|{}^{3}P^{\rm{o}}_{0}\right\rangle$ state are repumped to the $\left|{}^{1}S_{0}\right\rangle$ state, and the number of atoms in this state ($N_{\rm{e}}$) probed. Additionally, a third probe light pulse measures the background noise ($N_{\rm{b}}$). The excitation fraction ($p_{\rm{e}}$) is then determined by $(N_{\rm{e}}-N_{\rm{b}})/(N_{\rm{e}}+N_{\rm{g}}-2N_{\rm{b}})$ (see Appendix \ref{AP_a} for more detalis). The Rabi oscillation is measured by maintaining resonance on the transition and observing the maximum excitation fraction at various interrogation times. By taking into account the thermal distribution of the cold ensemble, the free-space Rabi frequency can be accurately extracted from the Rabi oscillation \cite{Blatt2009[20]}.
 
\begin{figure}
	\begin{center}
		\includegraphics[width=70mm]{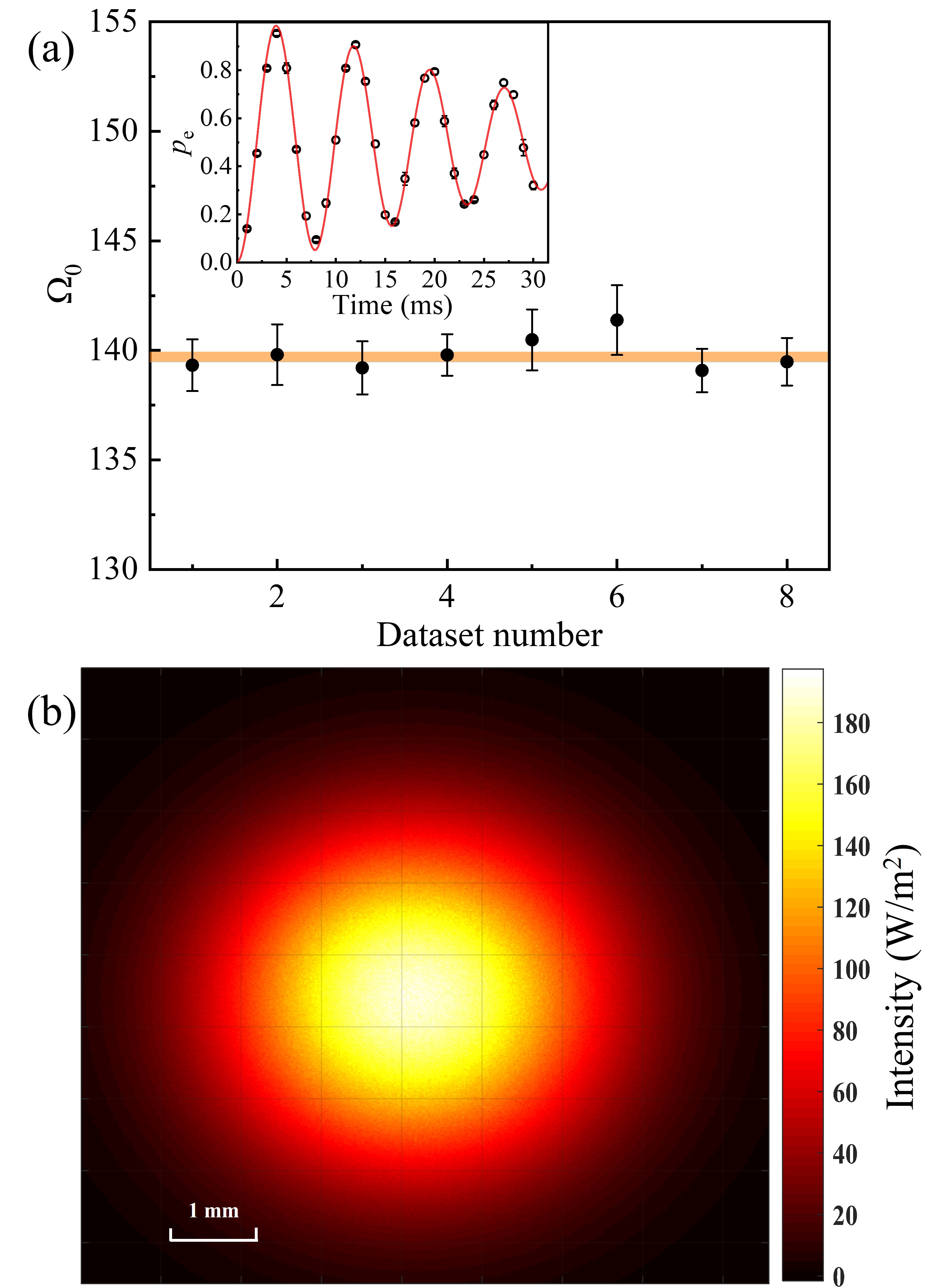}
		\caption{\label{fig3} Measurements of the free-space Rabi frequency $\Omega_{0}$ and laser intensity. (a) The measurements of $\Omega_{0}$. The orange area represents the average value and its corresponding $\pm1$ standard deviation. The inset shows a typical Rabi oscillation. (b) presents the reconstruction of the laser intensity. Its center is the expected intensity of interaction with atoms.}
	\end{center}
\end{figure}

\section{${}^{3}P^{\rm{o}}_{0}$ state lifetime measurement}
Due to the significantly larger size of the interrogation laser compared to the lattice, it is essential to secure the lattice at the center of the Gaussian interrogation laser beam. In this way, we can not only accurately ascertain the intensity of the laser's interaction with atoms but also significantly minimize the uncertainty in intensity attributed to errors in position measurement. To pinpoint the center position, we measure the Rabi frequency as a function of the position of the interrogation laser along the X and Y directions, as illustrated in Fig. \ref{fig2}. By adjusting the position of the translation stage, we initially determine the position of peak intensity in the X-direction ($X_{\rm{p}}$=6.69(5) mm). With the X scale fixed at $X_{\rm{p}}$, we proceed to measure the peak point in the Y-direction ($Y_{\rm{p}}$=7.13(4) mm). Gaussian fitting analysis reveals a $e^{-2}$ width of 6.77(16) mm for the X-direction and 6.66(22) mm for the Y-direction. Remarkably, these widths are consistent in the measurements obtained from the laser beam profiler (Thorlabs BP209-VIS/M) at the estimated lattice position, which recorded values of 6.87(7) mm and 6.74(7) mm for the X and Y directions, respectively. This agreement further validates our methodology and demonstrates the consistency of our experimental results.

Figure \ref{fig3}(a) shows measurements of the free-space Rabi frequency at (X=$X_{\rm{p}}$, Y=$Y_{\rm{p}}$). By analyzing the measured laser beam profile at the inferred distance of the lattice, we determine the ratio $R_{\rm{p}}$ of the measured total laser power to the total flux $C_{\rm{t}}$ recorded in the beam profile. Here $C_{\rm{t}}$ is the sum of the brightness level signals $C_{\rm{ij}}$ of all pixels in the digitally generated beam profile image. Thus, we can express the optical intensity of pixel (\textit{i,j}) as $R_{\rm{p}}C_{\rm{ij}}/A_{\rm{p}}$, where $A_{\rm{p}}$ denotes the pixel area.

As described in Ref. \cite{Lu2023[22]}, in terms of ${}^{87}$Sr, the hyperfine interaction can induce the $\left|{}^{3}P^{\rm{o}}_{1}\right\rangle$ and $\left|{}^{1}P^{\rm{o}}_{1}\right\rangle$ states mixing with the $\left|{}^{3}P^{\rm{o}}_{0}\right\rangle$ state, which opened the transition channel between the ground state and the $\left|{}^{3}P^{\rm{o}}_{0}\right\rangle$ state. According to Eq.(\ref{eq1}), the transition rate $ A (M^{\rm{g}}_{\rm{F}},M^{\rm{e}}_{\rm{F}})$ between the excited Zeeman state $\left| 5s5p \; {}^{3}P^{\rm{o}}_{0},F^{\rm{e}}=9/2,M^{\rm{e}}_{\rm{F}}=+9/2\right\rangle$ and the ground Zeeman state  $\left| 5s^{2} \; {}^{1}S_{0},F^{\rm{g}}=9/2,M^{\rm{g}}_{\rm{F}}=+9/2\right\rangle$ is \cite{Lu2023[22]}

\begin{equation}
\begin{split}
& A(M^{\rm{g}}_{\rm{F}},M^{\rm{e}}_{\rm{F}}) = \frac{2.02613\times 10^{18}}{\lambda^{\rm{3}}} \sum_{\rm{q}} \\
&|\left\langle "IJ^{\rm{g}}F^{\rm{g}}= IM^{\rm{g}}_{\rm{F}}"| O^{(1)}_{\rm{q}} |"IJ^{\rm{e}}F^{\rm{e}}= IM^{\rm{e}}_{\rm{F}}"\right\rangle |^{2},\label{eq3}
\end{split} 
\end{equation} 
where the state within the quotation marks describes the dominant component of the eigenvector and $ O^{(1)}_{\rm{q}}$ is the $q$th component of the electric dipole transition operator $O^{(1)}$. For the $\left| 5s5p \; {}^{3}P^{\rm{o}}_{0},F^{\rm{e}}=9/2, M^{\rm{e}}_{\rm{F}}=+9/2\right\rangle$ $\rightarrow$ $\left| 5s^{2} \; {}^{1}S_{0},F^{\rm{g}}=9/2,M^{\rm{g}}_{\rm{F}}=+9/2\right\rangle$ ($\rm{\pi}$ transition, $q=0$) in ${}^{87}$Sr, $\lambda=6984.457096$ $\rm{\AA}$ and the Rabi frequency is

\begin{equation}
\Omega_{\rm{M^{eg}_{F}}}= |\left\langle "IJ^{\rm{g}}F^{\rm{g}}= IM^{\rm{g}}_{\rm{F}}"| O^{(1)}_{\rm{q}} |"IJ^{\rm{e}}F^{\rm{e}}= IM^{\rm{e}}_{\rm{F}}"\right\rangle | \cdot \vec{E}/ \hbar. \label{eq4}
\end{equation} 

\begin{figure}
	\begin{center}
		\includegraphics[width=70mm]{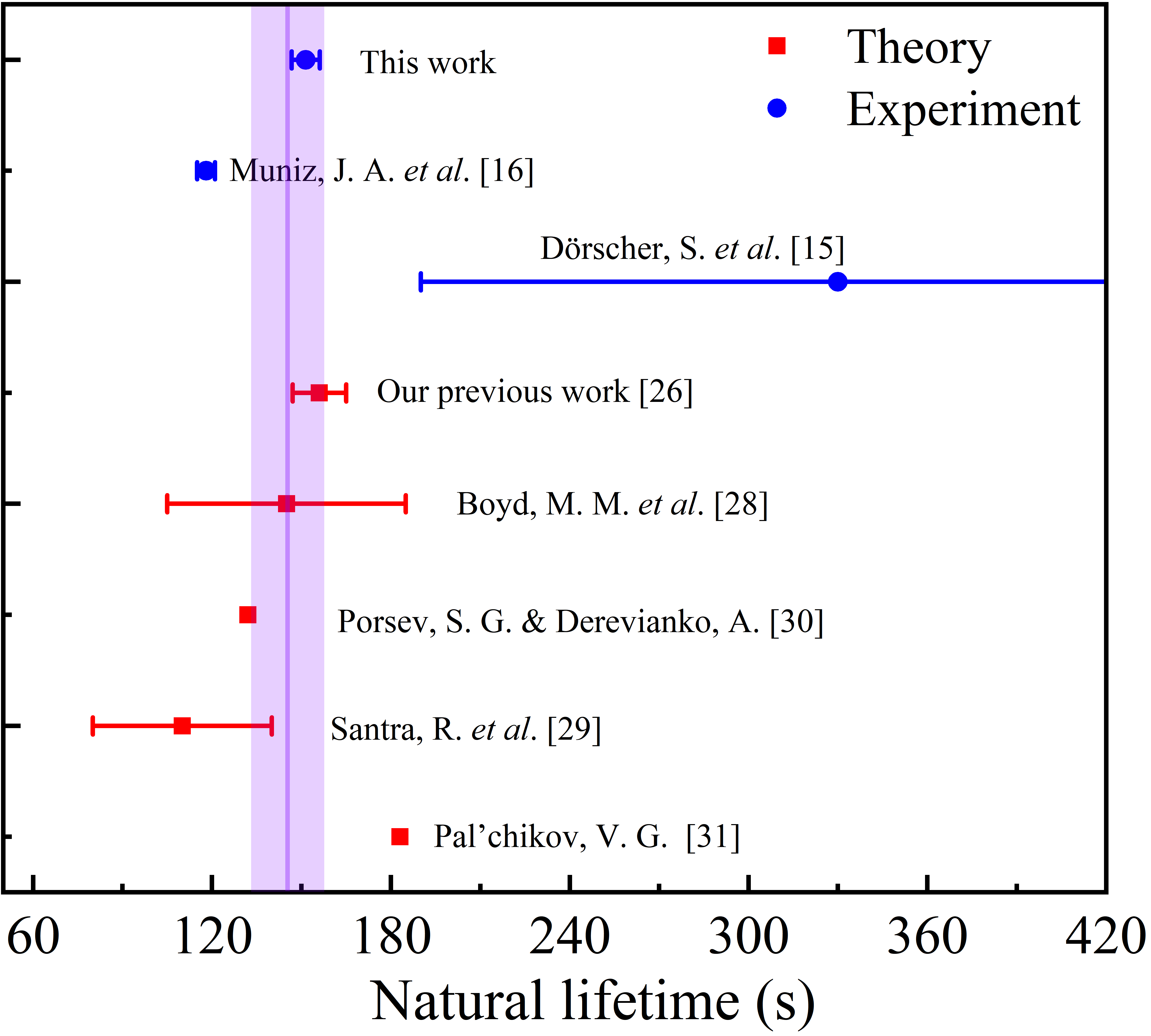}
		\caption{\label{fig4} Comparison of experimental (blue) and theoretical (red) results of the ${}^{87}$Sr state natural lifetime. Previous experimental results were based on rate equation analysis \cite{Dorscher2018[12]} and the cavity-enhanced dispersion measurements \cite{Muniz2021[13]}. Theoretical values are given in Refs. \cite{Lu2023[22],Boyd2007[24],Santra2004[25],Porsev2004[26],Palchikov2022[27]}. The purple line and shadow area show the arithmetic average value and the corresponding 1 standard error of theoretical results.}
	\end{center}
\end{figure}

In our experiment, taking into consideration the low atomic density (below $10^{12} \; \rm{cm^{-3}}$) and the spontaneous radiation rate of the $\left| 5s5p \; {}^{3}P^{\rm{o}}_{0}\right\rangle$ state, we can estimate that the value of $n_{\rm{r}}$ is approximately 1 with an error below $10^{-5}$ \cite{McCutcheon2022[add1]}. Additionally, given the ultra-low pressure of $10^{-9}$ Pa in the science chamber, we can safely estimate $n_{\rm{0}}$ to be 1. Combining Eq.(\ref{eq3}) and Eq. (\ref{eq4}), the transition rate of  $\left|{}^{3}P^{\rm{o}}_{0},F^{\rm{e}}=9/2, M^{\rm{e}}_{\rm{F}}=+9/2\right\rangle$ $\rightarrow$ $\left|{}^{1}S_{0},F^{\rm{g}}= 9/2,M^{\rm{g}}_{\rm{F}}=+9/2\right\rangle$ transition is experimentally determined as $5.41(17)\times10^{-3}$ $\rm{s^{-1}}$, and the corresponding dipole moment of $|\left\langle "IJ^{\rm{g}}F^{\rm{g}}= IM^{\rm{g}}_{\rm{F}}"| O^{(1)}_{\rm{0}} |"IJ^{\rm{e}}F^{\rm{e}}= IM^{\rm{e}}_{\rm{F}}"\right\rangle |$ is determined to be $3.0(1) \times 10^{-5}$ a.u. (a.u. indicates the atomic unit). There are two decaying channels from $\left|{}^{3}P^{\rm{o}}_{0},F^{\rm{e}}=9/2, M^{\rm{e}}_{\rm{F}}=+9/2\right\rangle$ to the ground state, including $\left|{}^{3}P^{\rm{o}}_{0},M^{\rm{e}}_{\rm{F}}=+9/2\right\rangle$ $\rightarrow$ $\left|{}^{1}S_{0},M^{\rm{g}}_{\rm{F}}=+9/2\right\rangle$ ($\rm{\pi}$ transition) and $\left|{}^{3}P^{\rm{o}}_{0},M^{\rm{e}}_{\rm{F}}=+9/2\right\rangle$ $\rightarrow$ $\left|{}^{1}S_{0},M^{\rm{g}}_{\rm{F}}=+7/2\right\rangle$ ($\rm{\sigma}$ transition), and the radiation rate ratio of $\rm{\pi}$ transition to $\rm{\sigma}$ transition is 4.5 [see Appendix \ref{AP_c} for details]. The total radiation rate of the $\left| 5s5p \; {}^{3}P^{\rm{o}}_{0},F^{\rm{e}}=9/2,M^{\rm{e}}_{\rm{F}}=+9/2\right\rangle$ level is determined to be $6.61(21)\times10^{-3} \; \rm{s^{-1}}$, corresponding to a natural lifetime of 151.4(48) s. In this experiment, the magnetic field strength is 0.4 G, resulting in a change of the Rabi frequency less than $5 \times 10^{-8}$ to the $ M^{\rm{e}}_{\rm{F}}=\pm9/2 $ Zeeman sub-levels \cite{Lu2023[22]}. Therefore, the magnetic field effect on the transition rate can be safely neglected. The natural lifetime of the $\left| 5s5p \; {}^{3}P^{\rm{o}}_{0}\right\rangle$ state, which accounts for the statistical average lifetime of all ten Zeeman sub-levels, is the same as that of the $\left| 5s5p \; {}^{3}P^{\rm{o}}_{0},F^{\rm{e}}=9/2,M^{\rm{e}}_{\rm{F}}=+9/2\right\rangle$ state.

\section{Correction and uncertainty evaluation for lifetime measurement}
The total uncertainty associated with the optical power meter (PM100D) and sensor (S130C) is 3\%, which translates to an uncertainty of 4.5 s in the lifetime measurement. The laser power interacting with atoms is determined by averaging the powers before and after passing through the windows. As a result, influences caused by the windows (such as reflection, absorption, and potential etalon effects) can be disregarded. The precision of reconstructing the laser intensity is limited by the knowledge of the precise position of the atoms. We can only measure the beam profile outside the science chamber at the calculated position. Measurement uncertainty will occur due to the size of the lattice and the positional inaccuracies. We have observed beam profiles deviating by $\pm40$ mm from the expected position and have determined that each millimeter of positional error contributes to a 0.188 s difference in the lifetime measurement. Taking into account the largest deviation of 8 mm (the radius of the cavity windows), this effect will contribute an uncertainty of 1.2 s to the lifetime measurement. The uncertainty of the Rabi frequency is determined to be 0.23\%, which corresponds to an uncertainty of 0.74 s. This uncertainty includes contributions from the precision of atomic temperature measurements (0.3\%) and the fitting error at 1 standard error (0.15\%). The center position of the laser beam is determined with an uncertainty of 0.22 mm as shown in Fig. \ref{fig2}, and the $e^{-2}$ waist diameter of the lattice is 0.1 mm. Therefore, we calculate the $\left|{}^{3}P^{\rm{o}}_{0}\right\rangle$ state lifetime using the average intensity around the measured center point with a radius of 0.3 mm. We conservatively estimate that this method may introduce a maximum error of 0.23 s based on the difference between the average and peak intensities.  

We observed that a small portion of the probe light extends beyond the beam profiler (with an effective diameter of 9 mm), resulting in a considerable systematic error. Through two-dimensional Gaussian fitting to deduce the complete intensity distribution, we determined that the power collected by the beam profiler is 1.81\% smaller than the total power. Additionally, the power meter sensor (with a diameter of 9.5 mm) will also leak a small amount of power, estimated at 1.17\%. Therefore, the systematic error due to the finite active aperture of the beam profiler and power meter is -0.97(52) s, with uncertainty estimated considering a perceived position error of the power meter of 0.5 mm. This systematic error can be suppressed by using an aperture close to the fiber collimator to restrict the size of the probe light. The presence of noise sidebands in the probe light introduces a systematic error as these sidebands contribute to the light power measured by the power meter but are not reflected in the Rabi oscillation. 

In the PDH scheme, the feedback loop tends to amplify noise beyond the feedback bandwidth, resulting in what is commonly referred to as 'servo-bumps' \cite{Akerman2015,Li2022,Chao2023}. These servo-bumps can lead to an overestimation of the measured light power since they do not contribute to the transition detection but instead occupy a small portion of the light power. By analyzing the beat signal between the probe light and a reference laser operating at 698 nm, we can determine the relative amplitudes of the servo-bumps in relation to the carrier signal. To prevent potential overlap of servo-bumps between the two lasers, the cavity transmission light of the reference laser is used to cancel out the servo-bumps \cite{Akerman2015}. Furthermore, we also observe spectral sidebands possibly created by a coherent source at $\pm19.2$ kHz and $\pm38.4$ kHz (these sidebands are also observed in the atomic response to an excitation of the 698 nm transition). Considering the contributions of both the servo-bumps and spectral sidebands, we estimate that they will introduce a correction of -1.24(26) s to the lifetime [see Appendix \ref{AP_d} for further details]. The uncertainty here reflects the variation in the calculated lifetime when we increase the PDH gain by 20\% from its typical operational point, where the amplitude of PDH error is minimized. The impact of servo-bumps can be strongly suppressed through techniques such as spectral filtering and post-correction methods \cite{Akerman2015,Li2022,Chao2023}.

Table \ref{tab:tab1} presents a summary of the evaluation results that contribute to the error in lifetime measurements. The total measurement uncertainty is determined to be 4.8 s (corresponding to a relative uncertainty of 3.2\%), which is mainly dominated by the power meter uncertainty. If the power meter uncertainty is reduced to 0.5\% \cite{Huntemann2016[23]} and a larger beam waist is used, this method has great potential to determine the lifetime with an uncertainty below 1\%. Figure \ref{fig4} summarizes the experimental and theoretical results of the $\left|{}^{3}P^{\rm{o}}_{0}\right\rangle$ state lifetime in ${}^{87}\rm{Sr}$. Our result not only agrees with our previous theoretical results using the multiconfiguration Dirac-Hartree-Fock method, but also accords with the arithmetic average value of all theoretical calculations. Thus, our result provides a clear point of reference for atomic structure calculations.

\begin{table}[ht]
\caption{\label{tab:tab1} Corrections of the lifetime measurement of the $5s5p$ ${}^{3}P^{\rm{o}}_{0}$ state.}
\begin{center}
\begin{ruledtabular}
\renewcommand{\arraystretch}{1}
\begin{tabular*}{8.5 cm}{lcc}
Source & Correction (s) & Uncertainty (s)\\ 
\hline
Optical power meter & - & 4.5 \\
Lattice position & - & 1.2 \\
Rabi frequency & - & 0.74 \\
Averaging optical intensity & - & 0.23 \\
Finite active aperture & -0.97 & 0.52 \\
servo-bumps and \\spectral sidebands & -1.24 & 0.26 \\
Total & -2.21 & 4.8\\

\end{tabular*}
\end{ruledtabular}
\end{center}
\end{table}

\section{CONCLUSION}
We use a straightforward yet effective technique for determining the lifetime of the $5s5p$ ${}^{3}P^{\rm{o}}_{0}$ in ${}^{87}\rm{Sr}$. By combining measurements of the free-space Rabi frequency and reconstructing the laser intensity to determine the transition matrix element, we infer the lifetime to be 151.4(48) seconds. This method explained here holds potential for broad application in other atomic species, including other alkaline-earth atoms and ions \cite{Dube2013[28],Barrett2019[29],Huang2022[30],McGrew2018[31],Filzinger2023[32]}, ${}^{175}\rm{Lu}^{+}$\cite{Zhang2023[33]}, $\rm{Pb}^{2+}$ \cite{Beloy2021[34]} and highly charged ions \cite{Steven2022[35]}. Accurate lifetime measurements contribute to a deeper understanding of the atomic structure, enable testing of theoretical models, and support quantitative experimental investigations of the hyperfine quenching effect \cite{Johnson2010[36]} by measuring the lifetimes of all stable isotopes (such as in Yb \cite{Roman2021[37]}). Additionally, the precise determination of laser intensity can also have a positive impact on other measurements, such as the measurement of the differential polarizability \cite{Huntemann2016[23]}.  \\

\section{ACKNOWLEDGMENTS}
We thank Lin-Xiang He from the Innovation Academy for Precision Measurement Science and Technology, CAS for helpful discussions, and Luis A. Orozco from the University of Maryland and Yan-Ting Zhao from Shanxi University for his careful reading of the manuscript. This work is supported by the National Natural Science Foundation of China (Grant No. 12203057), and the Strategic Priority Research Program of the Chinese Academy of Sciences (Grant No. XDB35010202).\\

\appendix
\section{Preparation of the cold ${}^{87}$Sr ensemble}\label{AP_a}
\begin{figure}
	\begin{center}
		\includegraphics[width=70mm]{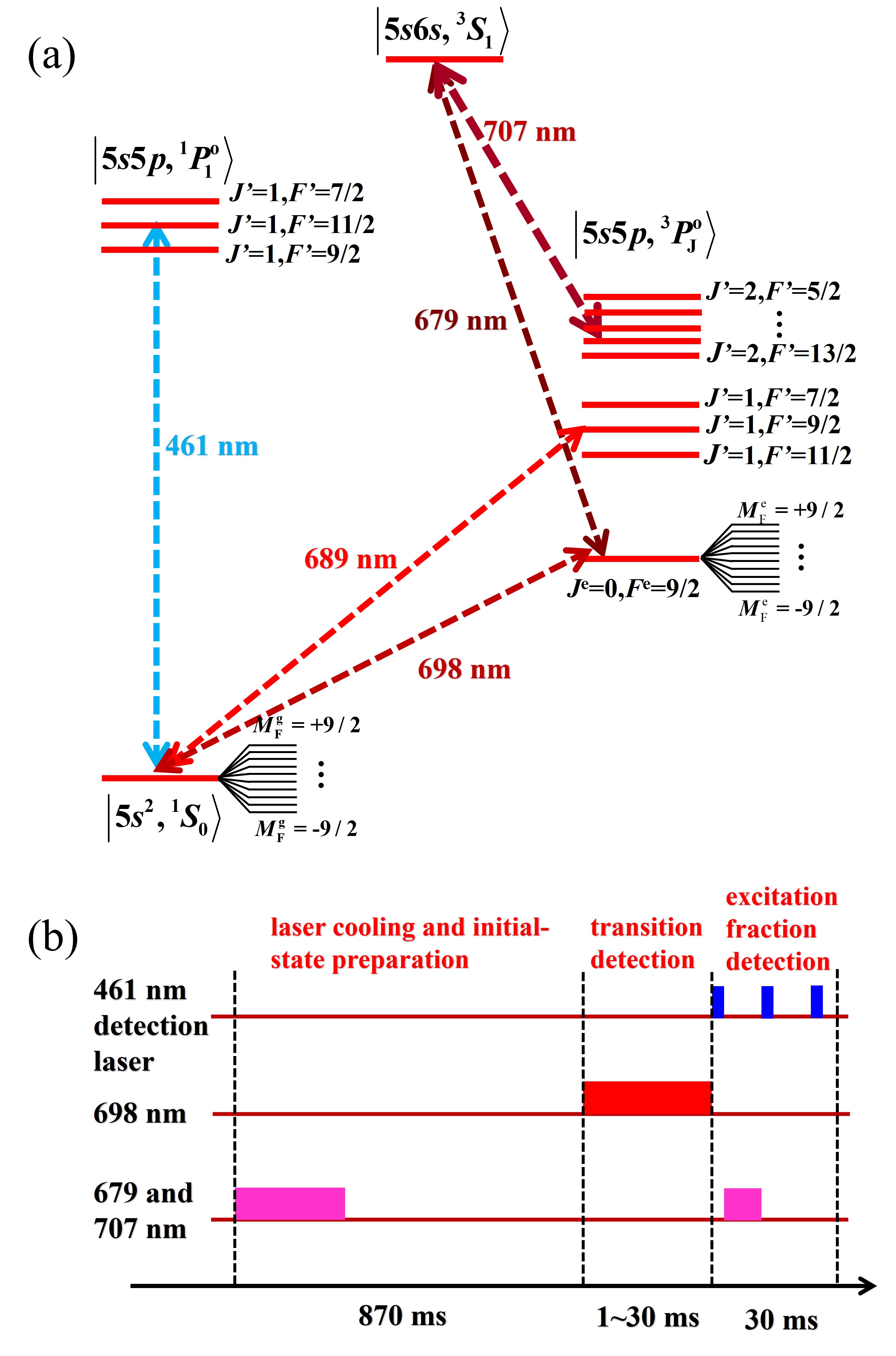}
		\caption{\label{fig5} (a) Energy levels of ${}^{87}$Sr. (b) Timing scheme of excitation fraction detection.}
	\end{center}
\end{figure}

Figure \ref{fig5}(a) shows the energy levels of ${}^{87}$Sr relevant to this experiment. After preliminary cooling by passing the Zeeman slower, atoms from atomic oven are trapped and cooled by the blue magneto-optical trap (MOT). The bule MOT is based on the $\left|{}^{1}P^{\rm{o}}_{1}, F\rm{'}=11/2\right\rangle$ $\rightarrow$ $\left|{}^{1}S_{0}, F^{\rm{g}}=9/2\right\rangle$ transition and can reduce atomic temperature to $1\sim5$ mK. The $\left|{}^{3}P^{\rm{o}}_{0}\right\rangle$ $\rightarrow$ $\left|{}^{3}S_{1}\right\rangle$ and $\left|{}^{3}P^{\rm{o}}_{2}\right\rangle$ $\rightarrow$ $\left|{}^{3}S_{1}\right\rangle$ transitions can repump atoms populated in ${}^{3}P^{\rm{o}}_{0}$ and ${}^{3}P^{\rm{o}}_{2}$ states back to $^1S_0$ state, increasing the number of trapped atoms. Following the bule MOT, the $\left|{}^{3}P^{\rm{o}}_{1}, F\rm{'}=11/2\right\rangle$ $\rightarrow$ $\left|{}^{1}S_{0}, F^{\rm{g}}=9/2\right\rangle$ and $\left|{}^{3}P^{\rm{o}}_{1}, F\rm{'}=9/2\right\rangle$ $\rightarrow$ $\left|{}^{1}S_{0}, F^{\rm{g}}=9/2\right\rangle$ transitions are simultaneously used to further decrease the thermodynamic temperature of atoms to 3 $\rm{\mu K}$. Then, atoms are loaded into the optical lattice by turning off MOT lasers and quadrupole magnetic field. The lattice laser frequency of 368 554 485.0(1) MHz is stabilized to an ultra-low-expansion (ULE) cavity with a fineness of 20,000. The lattice AC Stark shift is almost zero at this "magic frequency". Before the energy filtering process, atoms are optically repumped into the $\left|{}^{1}S_{0},F^{\rm{g}}=9/2, M^{\rm{g}}_{\rm{F}}=+9/2\right\rangle$ state using a right-handed circularly polarized light resonant on the $\left|{}^{3}P^{\rm{o}}_{1}, F^{\rm{e}}=9/2\right\rangle$ $\rightarrow$ $\left|{}^{1}S_{0}, F^{\rm{g}}=9/2\right\rangle$ transition.

Figure \ref{fig5}(b) illustrates the timing scheme for excitation fraction detection. Following the interrogation of the 698 nm transition, three 461 nm pulses (each with a duration of 2 ms) are employed to determine the excitation fraction. The first 461 nm pulse is used to ascertain the number of atoms remaining in the $\left|{}^{1}S_{0}\right\rangle$ state. Strong radiation pressure compels atoms out of the lattice. Prior to applying the second 461 nm pulse to measure atoms in the $\left|{}^{3}P_{0}\right\rangle$ state, 679 nm and 707 nm lasers are used (with pulse durations of 10 ms) to repump atoms populated in the $\left|{}^{3}P_{0}\right\rangle$ state back to the $\left|{}^{1}S_{0}\right\rangle$ state. The third 461 nm pulse is utilized for assessing background noise.  

\section{Determination of the free-space Rabi frequency}\label{AP_b}
As atoms have the Boltzmann distribution in the external states, the excitation fraction with Rabi detection at zero frequency detuning can be expressed by \cite{Blatt2009[20]}
\begin{equation}
\begin{split}
P_{carrier}(t_{\rm{p}}) &= \sum_{\rm{n_{x}=1}}^{\rm{N_{x}}} \sum_{\rm{n_{z}=1}}^{\rm{N_{z}}} Q_{\rm{n_{x}}}(T_{\rm{r}}) \\
&Q_{\rm{n_{z}}}(T_{\rm{z}}) \sin^{2}(0.5t_{\rm{p}}  \Omega_{\rm{n_{x}},\rm{n_{z}}}).\label{eqA1}
\end{split} 
\end{equation} 

In Eq.(\ref{eqA1}), $t_{\rm{p}}$ indicates the interrogation time. $\Omega_{\rm{n_{x}},\rm{n_{z}}}=\Omega_{0} e^{-\eta^{2}_{\rm{x}}/2} e^{-\eta^{2}_{\rm{z}}/2} L_{\rm{n_{x}}}(\rm{\eta^{2}_{x}}) \it{L}_{\rm{n_{z}}}(\rm{\eta^{2}_{z}})$ represents the effective Rabi frequency in the quantum number of external states ($\rm{n_{x}}$, $\rm{n_{z}}$), where $L_{\rm{n}}$ is the $n$th order Laguerre polynomial, and $\eta_{\rm{x}}=\delta \theta / \lambda_{\rm{p}} \sqrt{h/2m_{\rm{a}} v_{\rm{r}} } $ and $\eta_{\rm{z}}= 1 / \lambda_{\rm{p}} \sqrt{h/2m_{\rm{a}} v_{\rm{z}} } $ are the Lamb-Dick parameters, in the transverse and longitudinal directions, respectively. Herein, $\lambda_{\rm{p}}$ is the wavelength of interrogation, $v_{\rm{z}}$ is the longitudinal trap frequency, $v_{\rm{r}}$ indicates the transverse trap frequency, $m_{\rm{a}}$ is the mass of ${}^{87}$Sr and $\delta \theta $ represents effective misalignment between lattice and interrogation laser beams. $Q_{\rm{n_{x}}}(T_{\rm{r}})=(1-e^{-hv_{\rm{r}} /k_{\rm{b}} T_{\rm{r}} })e^{-n_{\rm{x}} h v_{\rm{r}} /k_{\rm{b}} T_{\rm{r}} }$ and $Q_{\rm{n_{z}}}(T_{\rm{z}})=(1-e^{-hv_{\rm{z}} /k_{\rm{b}} T_{\rm{z}} })e^{-n_{\rm{z}} h v_{\rm{z}} /k_{\rm{b}} T_{\rm{z}} }$ are the normalized Boltzmann weights. $N_{\rm{x}}$ and $N_{\rm{z}}$ are the number of states in the trap at the transverse and longitudinal directions, respectively. According to the resolved sideband spectrum in longitudinal and transverse directions, parameters of $v_{\rm{z}}$=91.44 kHz, $v_{\rm{r}}$= 340 Hz, $T_{\rm{z}}$= 2.7(3) $\rm{\mu K}$, $T_{\rm{r}}$= 3.9(4) $\rm{\mu K}$, $N_{\rm{z}}$=7, $N_{\rm{x}}$=1787 can be determined. The free-space Rabi frequency $\Omega_{0}$ can be extracted by fitting the data of Rabi oscillation using Eq. (\ref{eqA1}) with free parameters of $\delta \theta $ and $\Omega_{0}$. \\

\section{The transition rate ratio of $\rm{\pi}$ transition to $\rm{\sigma}$ transition}\label{AP_c}
In ${}^{87}$Sr, the hyperfine interaction can induce the ${}^{3}P^{\rm{o}}_{1} $ and ${}^{1}P^{\rm{o}}_{1} $ states mixing with the ${}^{3}P^{\rm{o}}_{0} $ state, which opened the transition channel between the ground state and the ${}^{3}P^{\rm{o}}_{0} $ state. According to the theory of the unexpected transition \cite{Brage2009,Andersson2009,Li2011,Grumer2014[16]}, the wave function of the ${}^{3}P^{\rm{o}}_{0} $ state can be written as linear combination of the ${}^{3}P^{\rm{o}}_{1} $ and ${}^{1}P^{\rm{o}}_{1} $ perturbing states \cite{Lu2023[22]}
\begin{equation}
\begin{split}
&\left| "5s5p \; {}^{3}P^{\rm{o}}_{0}, F^{\rm{e}} = IM^{\rm{e}}_{\rm{F}}"\right\rangle = \\
& \left| 5s5p \; {}^{3}P^{\rm{o}}_{0}, F^{\rm{e}}= IM^{\rm{e}}_{\rm{F}}\right\rangle + \\
&\sum_{\rm{s=1,3}} d_{\rm{s}} \left| "5s5p \; {}^{\rm{s}}P^{\rm{o}}_{1}, F'= IM^{\rm{e}}_{\rm{F}}"\right\rangle, \label{eqA2}
\end{split} 
\end{equation}
where the state within the quotation marks describes the dominant component of the eigenvector. Other interactions from different configurations can be neglected due to their large energy separations and weak hyperfine interactions. In first-order perturbation theory, the mixing coefficients are given by
\begin{equation}
d_{\rm{s}}=\frac {\left\langle 5s5p \; {}^{\rm{s}}P_{1} F' M_{\rm{F}}| H_{\rm{hfs}} | 5s5p \; {}^{\rm{3}}P_{0} F M_{\rm{F}} \right\rangle } {E(5s5p \; {}^{\rm{3}}P_{0} F M_{\rm{F}})-E(5s5p \; {}^{\rm{s}}P_{1} F' M_{\rm{F}})}. \label{eqA3}
\end{equation}

The ground state is separate from other states, its wave function is given as
\begin{equation}
\left| "5s^{2} \; {}^{1}S_{0}, F^{\rm{g}} = IM^{\rm{g}}_{\rm{F}}"\right\rangle = \left| 5s^{2} \; {}^{1}S_{0}, F^{\rm{g}} = IM^{\rm{g}}_{\rm{F}}\right\rangle. \label{eqA4}
\end{equation}

Therefore, the hyperfine induced transition rate $A(M^{\rm{g}}_{\rm{F}},M^{\rm{e}}_{\rm{F}})$ between the excited Zeeman state $\left| 5s5p \; {}^{3}P^{\rm{o}}_{0}, F^{\rm{e}}= IM^{\rm{e}}_{\rm{F}} \right\rangle$ and the ground Zeeman state $\left| 5s^{2} \; {}^{1}S_{0}, F^{\rm{g}}= IM^{\rm{g}}_{\rm{F}} \right\rangle$ is presented as
\begin{equation}
\begin{split}
& A(M^{\rm{g}}_{\rm{F}},M^{\rm{e}}_{\rm{F}}) = \frac{2.02613\times 10^{18}}{\lambda^{\rm{3}}} \sum_{\rm{q}} \\
&|\left\langle "5s^{2} {}^{1}S_{0}\; , F^{\rm{g}}= IM^{\rm{g}}_{\rm{F}}"| O^{(1)}_{\rm{q}} |"5s5p \; {}^{3}P^{\rm{o}}_{0} , F^{\rm{e}}= IM^{\rm{e}}_{\rm{F}}"\right\rangle |^{2}.\label{eqA5}
\end{split} 
\end{equation} 

The \textbf{E}1 transition matrix element is in the atomic unit (a.u.), but the units of $\Omega_{\rm{M^{eg}_{F}}}$ and $I_{0}$ are Hz and $\rm{W/m^{2}}$, respectively. Thus, the measured value of the transition matrix element should be divided by $4 \rm{\pi} \varepsilon_{0}\hbar^{2}/m_{\rm{e}} e_{0} $ for conversion to the atomic unit, where $m_{\rm{e}}$ represents the mass of the electron, and $e_{0}$ represents the charge of the electron. By submitting Eqs. (\ref{eqA2}) and (\ref{eqA4}) into the above equation, the transition rate $A(M^{\rm{g}}_{\rm{F}},M^{\rm{e}}_{\rm{F}})$ is then given as
\begin{equation}
\begin{split}
& A(M^{\rm{g}}_{\rm{F}},M^{\rm{e}}_{\rm{F}}) = \frac{2.02613\times 10^{18}}{\lambda^{\rm{3}}} \times \\
&\sum_{\rm{q}}| \sum_{\rm{s=1,3}} d_{\rm{s}} \sqrt{2F^{\rm{g}}+1} \sqrt{2F^{\rm{e}}+1} \\
&{\begin{pmatrix}
    F^g   & 1 & F^{\rm{e}}\\
  - M^g_F & q & M^e_F
\end{pmatrix}} 
{\begin{Bmatrix}
    J^g   & F^g   & I\\
    F^{\rm{e}}    & J^e   & 1
\end{Bmatrix}
\langle {5s^2\ ^1\!S_0} || O^{(1)}_{\rm{q}} || {5s5p \; {}^{\rm{s}}\!P^{\rm{o}}_{1}}\rangle{|^2}}.\label{eqA6}
\end{split} 
\end{equation} 

According to Eq. (\ref{eqA6}), the ratio of the \textbf{E}1 transition matrix elements between the $\left|5s5p \; {}^{3}P^{\rm{o}}_{0},F^{\rm{e}}= 9/2,M^{\rm{e}}_{\rm{F}}=+9/2\right\rangle$ $\rightarrow$ $\left|5s^{2} \;  {}^{1}S_{0},F^{\rm{g}}= 9/2,M^{\rm{g}}_{\rm{F}}=+9/2\right\rangle$ and $\left|5s5p \; {}^{3}P^{\rm{o}}_{0},F^{\rm{e}}= 9/2,M^{\rm{e}}_{\rm{F}}=+9/2\right\rangle$ $\rightarrow$ $\left|5s^{2} \; {}^{1}S_{0},F^{\rm{g}}= 9/2,M^{\rm{g}}_{\rm{F}}=+7/2\right\rangle$ transitions is $ {\begin{pmatrix}
    9/2   & 1   & 9/2\\
    -9/2    & 0   & 9/2
\end{pmatrix}}/{\begin{pmatrix}
    9/2   & 1   & 9/2\\
    -7/2    & -1   & 9/2 
\end{pmatrix}}=-3 / \sqrt{2}$. The corresponging ratio of transition rate is its square and equals 9/2.\\ 

\section{Measurements of the servo-bumps and spectral peaks}\label{AP_d} 

\begin{figure}[ht]
	\begin{center}
		\includegraphics[width=70mm]{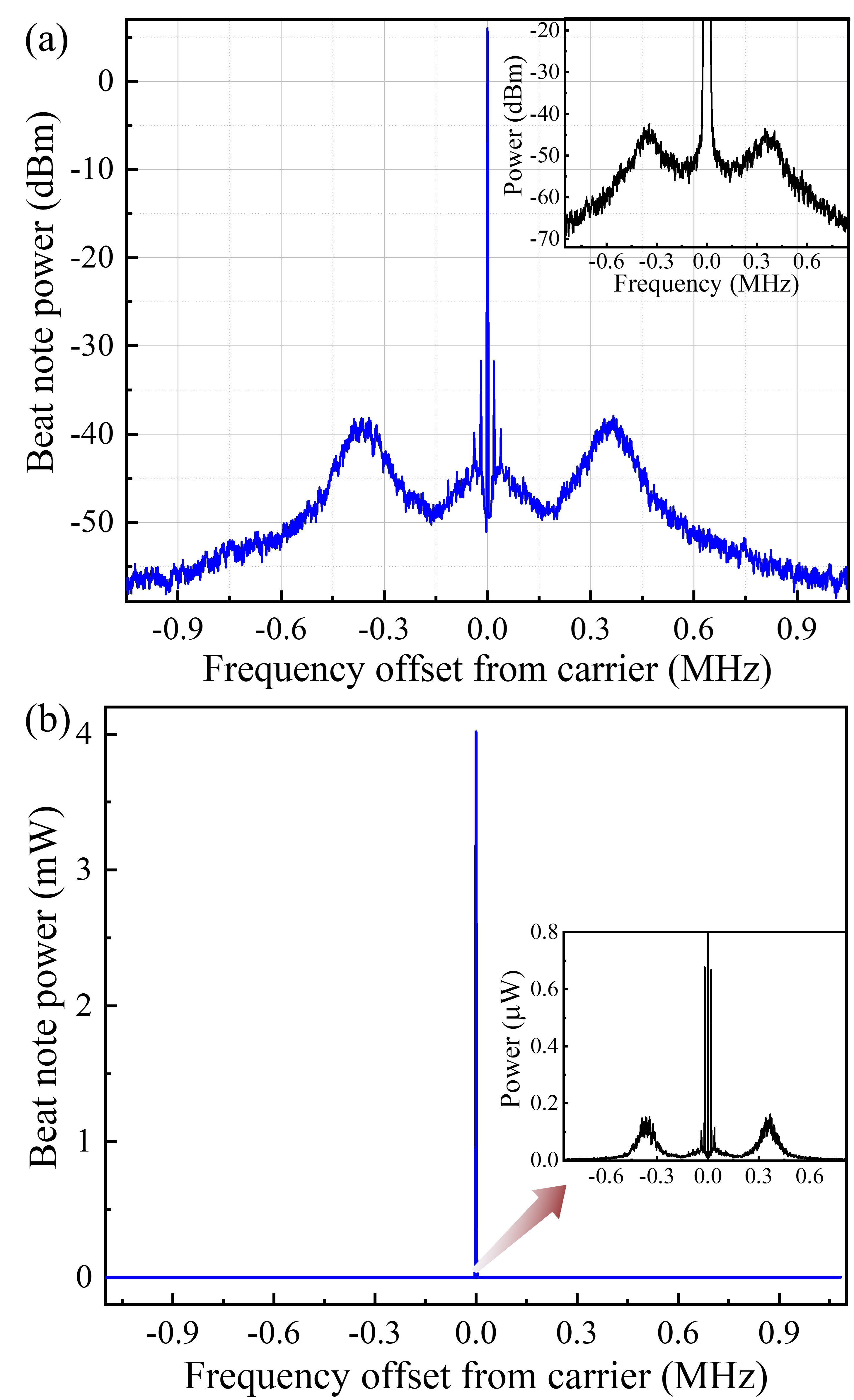}
		\caption{\label{fig6} Spectra of the 698 nm probe laser. (a) The servo-bumps and spectral peaks within the logarithmic Y-axis. The inset shows the Fourier analysis of the PDH error signal of the probe light used in this work. (b) The same data in (a) but shown in the linear Y-axis. The inset shows the details of the small servo-bumps and spectral peaks.}
	\end{center}
\end{figure}

The probe laser utilized in this study is compared with another 698 nm reference laser. The reference laser employs the cavity transmission light (with a power of approximately 19.8 $\rm{\mu W}$) to filter its servo-bumps. A photodetector (FPD310-FS-VIS, MenloSystems) with a frequency response range from 1 MHz to 1.5 GHz is employed to detect the beat signal. In Figure \ref{fig6}(a), the power spectral densities of the beat signal are displayed through Fourier analysis using a spectrum analyzer (N9030A, KEYSIGHT). By simultaneously examining the Fourier analysis of the PDH error, we validate that the observed servo-bumps originate from the laser used in this experiment \cite{Akerman2015}. In both cases, the peak frequencies of the servo-bumps are around $\pm 0.352(4)$ MHz. To quantify the contributions of the servo-bumps and spectral peaks to the total light power, we integrate the regions excluding the carrier in the linear coordinates presented in Fig. \ref{fig6}(b). It is estimated that the servo-bumps and spectral peaks contribute 0.815(8)\% to the total power, with the uncertainty indicating the 1$\sigma$ standard error of 10 measurements.

\begin{thebibliography}{100}

\bibitem{Bothwell2022[1]}
\href{ https://doi.org/10.1038/s41586-021-04349-7} { T. Bothwell, C. J. Kennedy, A. Aeppli, D. Kedar, J. M. Robinson, E. Oelker, A. Staron, and J. Ye, Nature \textbf{602}, 420 (2022).}

\bibitem{Sanner2019[2]}
\href{https://doi.org/10.1038/s41586-019-0972-2} { C. Sanner, N. Huntemann, R. Lange, C. Tamm, E. Peik, M. S. Safronova, and S. G. Porsev, Nature \textbf{567}, 204, (2019).}

\bibitem{Takamoto2022[3]}
\href{https://doi.org/10.1063/5.0087894 } { M. Takamoto, Y. Tanaka, and H. Katori, Appl. Phys. Lett. \textbf{120}, 140502 (2022).}


\bibitem{Drscher2021[4]}
\href{https://doi.org/ 10.1088/1681-7575/abc86f } { S. D$\rm{\ddot{o}}$rscher, N. Huntemann, R. Schwarz, R. Lange, E. Benkler, B. Lipphardt, U. Sterr, E. Peik, and C. Lisdat, Metrologia \textbf{58}, 015005 (2021). }


\bibitem{Golovizin2019[5]}
\href{https://doi.org/10.1038/s41467-019-09706-9} {A. Golovizin, E. Fedorova, D. Tregubov, D. Sukachev, K. Khabarova, V. Sorokin, and N. Kolachevsky, Nat. Commun. \textbf{10}, 1724 (2019).}


\bibitem{Grotti2018[6]}
\href{https://doi.org/10.1038/s41567-017-0042-3} { J. Grotti, S. Koller, S. Vogt, \textit{et al}. Nat. Phys. \textbf{14}, 437 (2018).}


\bibitem{Lu2021[7]}
\href{https://doi.org/10.1103/PhysRevLett.127.033601} { X. T. Lu, T. Wang, T. Li, C. H. Zhou, M. J. Yin, Y. B. Wang, X. F. Zhang, and H. Chang, Phys. Rev. Lett. \textbf{127}, 033601 (2021).}

\bibitem{Daley2008[8]}
\href{https://doi.org/10.1103/PhysRevLett.101.170504} { A. J. Daley, M. M. Boyd, J. Ye, and P. Zoller, Phys. Rev. Lett. \textbf{101}, 170504 (2008).}


\bibitem{McGrew2018[31]}
\href{https://doi.org/10.1038/s41586-018-0738-2} { W. F. McGrew, X. Zhang, R. J. Fasano, S. A. Sch$\rm{\ddot{a}}$ffer, K. Beloy, D. Nicolodi, R. C. Brown, N. Hinkley, G. Milani, M. Schioppo, T. H. Yoon, and A. D. Ludlow, Nature \textbf{564}, 87 (2018).}


\bibitem{Porsev2006}
\href{https://doi.org/10.1103/PhysRevA.74.020502} {S. G. Porsev, and A. Derevianko. Phys. Rev. A \textbf{74}, 020502(R) (2006).}

\bibitem{Safronova2013}
\href{https://doi.org/10.1103/PhysRevA.87.012509} {M. S. Safronova, S. G. Porsev, U. I. Safronova, M. G. Kozlov, and Charles W. Clark. Phys. Rev. A \textbf{87}, 012509 (2013).}


\bibitem{Beloy2012[9]}
\href{https://doi.org/10.1103/PhysRevA.86.051404} { K. Beloy, J. A. Sherman, N. D. Lemke, N. Hinkley, C. W. Oates, and A. D. Ludlow, Phys. Rev. A \textbf{86}, 051404(R) (2012).}


\bibitem{Olmschenk2009[10]}
\href{https://doi.org/10.1103/PhysRevA.80.022502} { S. Olmschenk, D. Hayes, D. N. Matsukevich, P. Maunz, D. L. Moehring, K. C. Younge, and C. Monroe, Phys. Rev. A \textbf{80}, 022502 (2009).}


\bibitem{Yasuda2004[11]}
\href{https://doi.org/10.1103/PhysRevLett.92.153004} {M. Yasuda, and H. Katori, Phys. Rev. Lett. \textbf{92}, 153004 (2004).}


\bibitem{Dorscher2018[12]}
\href{https://doi.org/10.1103/PhysRevA.97.063419} {S. D$\rm{\ddot{o}}$rscher, R. Schwarz, A. Al-Masoudi, S. Falke, U. Sterr, and C. Lisdat, Phys. Rev. A \textbf{97}, 063419 (2018).}


\bibitem{Muniz2021[13]}
\href{https://doi.org/10.1103/PhysRevResearch.3.023152} { J. A. Muniz, D. J. Young, J. R. K. Cline, and J. K. Thompson, Phys. Rev. Res. \textbf{3}, 023152 (2021).}


\bibitem{Hettrich2015[14]}
\href{https://doi.org/10.1103/PhysRevLett.115.143003} { M. Hettrich, T. Ruster, H. Kaufmann, C. F. Roos, C. T. Schmiegelow, F. Schmidt-Kaler, and U. G. Poschinger, Phys. Rev. Lett. \textbf{115}, 143003 (2015).}

\bibitem{Lange2021[15]}
\href{https://doi.org/10.1103/PhysRevLett.127.213001} { R. Lange, A. A. Peshkov, N. Huntemann, Chr. Tamm, A. Surzhykov, and E. Peik, Phys. Rev. Lett. \textbf{127}, 213001 (2021).}


\bibitem{Huntemann2016[23]}
\href{https://doi.org/10.1103/PhysRevLett.116.063001} { N. Huntemann, C. Sanner, B. Lipphardt, Chr. Tamm, and E. Peik, Phys. Rev. Lett. \textbf{116}, 063001 (2016).}


\bibitem{Grumer2014[16]}
\href{https://doi.org/10.1088/0031-8949/89/11/114002} { J. Grumer, T. Brage, M. Andersson, J. G. Li, P. J$\rm{\ddot{o}}$nsson, W. X. Li, Y. Yang, R. Hutton, and Y. M. Zou, Phys. Scr. \textbf{89}, 114002 (2014).}


\bibitem{Scar1981[17]}
\href{https://doi.org/10.1103/PHYSREVA.24.883} { D. Scar, Lloyd A. Hackel, Michael A. Johnson, and Michael C. Rushford, Phys. Rev. A \textbf{24}, 883 (1981). }


\bibitem{Kong2020[18]}
\href{https://doi.org/10.1088/1674-1056/ab9290} { D. H. Kong, Z. H. Wang, F. Guo, Q. Zhang, X. T. Lu, Y. B. Wang, and H. Chang, Chin. Phys. B \textbf{29}, 070602 (2020).}


\bibitem{Kong2021[19]}
\href{https://doi.org/10.7498/aps.70.20201204} { D. H. Kong, F. Guo, T. Li, X. T. Lu, Y. B. Wang, and H. Chang, Acta  Phys.  Sin. \textbf{70}, 030601 (2021).}


\bibitem{Blatt2009[20]}
\href{https://doi.org/10.1103/PhysRevA.80.052703} { S. Blatt, J. W. Thomsen, G. K. Campbell, A. D. Ludlow, M. D. Swallows, M. J. Martin, M. M. Boyd, and J. Ye, Phys. Rev. A \textbf{80}, 052703 (2009).}

\bibitem{Nagourney1986[21]}
\href{https://doi.org/10.1103/PhysRevLett.56.2797} {W. Nagourney, I. Sandberg, and H. Dehmelt, Phys. Rev. Lett. \textbf{56}, 2797 (1986)}


\bibitem{Lu2023[22]}
\href{https://doi.org/10.1103/PhysRevA.108.012820} { X. T. Lu, F. Guo, Y. B. Wang, M. Feng, T. Liang, B. Q. Lu, and Hong Chang. Phys. Rev. A \textbf{108}, 012820 (2023).}


\bibitem{McCutcheon2022[add1]}
\href{https://doi.org/10.1016/j.optcom.2021.127583} { R. A. McCutcheon, and S. F. Yelin. Opt. Commun. \textbf{505}, 127583 (2022).}


\bibitem{Boyd2007[24]}
\href{https://doi.org/10.1103/PhysRevA.76.022510} { M. M. Boyd, T. Zelevinsky, A. D. Ludlow, S. Blatt, T. Zanon-Willette, S. M. Foreman, and J. Ye, Phys. Rev. A \textbf{76}, 022510 (2007).}


\bibitem{Santra2004[25]}
\href{https://doi.org/10.1103/PhysRevA.69.042510} { R. Santra, K. V. Christ, and C. H. Greene, Phys. Rev. A \textbf{69}, 042510 (2004).}


\bibitem{Porsev2004[26]}
\href{https://doi.org/10.1103/PhysRevA.69.042506} { S. G. Porsev, and A. Derevianko, Phys. Rev. A \textbf{69}, 042506 (2004).}

\bibitem{Palchikov2022[27]}
\href{https://www.eftf.org/fileadmin/conferences/eftf/documents/Proceedings/proceedingsEFTF2002.zip} { Pal'chikov, V. G. 16th European Frequency and Time Forum (St. Petersburg State University of Aerospace Instrumentation, St. Petersburg, Rosja, 2002), p. E-002.}

\bibitem{Akerman2015}
\href{https://doi.org/10.1088/1367-2630/17/11/113060} { N. Akerman, N. Navon, S. Kotler, Y. Glickman, and R. Ozeri, New J. Phys. \textbf{17}, 113060 (2015)}

\bibitem{Li2022}
\href{https://doi.org/10.1103/PhysRevApplied.18.064005} { L. T. Li, W. Huie , N. Chen, B. DeMarco, and J. P. Covey, Phys. Rev. Applied \textbf{18}, 064005 (2022)}

\bibitem{Chao2023}
\href{https://arxiv.org/pdf/2309.09759.pdf} { Y. X. Chao, Z. X. Hua, X. H. Liang, Z. P. Yue, L. You, and M. K. Tey, arXiv:2309.09759v1 (2023)}

\bibitem{Dube2013[28]}
\href{https://doi.org/10.1103/PhysRevA.87.023806} { P. Dub$\rm{\acute{e}}$, A. A. Madej, Z. C. Zhou, and John E. Bernard, Phys. Rev. A \textbf{87}, 023806 (2013).}

\bibitem{Barrett2019[29]}
\href{https://doi.org/10.1103/PhysRevA.100.043418} { M. D. Barrett, K. J. Arnold, and M. S. Safronova, Phys. Rev. A \textbf{100}, 043418 (2019)}


\bibitem{Huang2022[30]}
\href{https://doi.org/10.1103/PhysRevApplied.17.034041} { Y. Huang, B. L. Zhang, M. Y. Zeng, Y. M. Hao, Z. X. Ma, H. Q. Zhang, H. Guan, Z. Chen, M. Wang, and K. L. Gao, Phys. Rev. Applied \textbf{17}, 034041 (2022).}


\bibitem{Filzinger2023[32]}
\href{https://doi.org/10.1103/PhysRevLett.130.253001} { M. Filzinger, S. D$\rm{\ddot{o}}$rscher, R. Lange, J. Klose, M. Steinel, E. Benkler, E. Peik, C. Lisdat, and N. Huntemann, Phys. Rev. Lett. \textbf{130}, 253001 (2023).}


\bibitem{Zhang2023[33]}
\href{https://doi.org/10.1126/sciadv.adg197} {Z. Q. Zhang, J. A. Kyle, K. Rattakorn, and D. B. Murray, Sci. Adv. \textbf{9}, eadg1971 (2023).}


\bibitem{Beloy2021[34]}
\href{https://doi.org/10.1103/PhysRevLett.127.013201} { K. Beloy, Phys. Rev. Lett. \textbf{127}, 013201 (2021). }


\bibitem{Steven2022[35]}
\href{https://doi.org/10.1038/s41586-022-05245-4} { A. K. Steven, J. S. Lukas, M. Peter, \textit{et al}. Nature \textbf{611}, 43 (2022).}


\bibitem{Johnson2010[36]}
\href{https://doi.org/10.1139/p11-018} {W. R. Johnson, Can. J. Phys. \textbf{99}, 1 (2010).}

\bibitem{Roman2021[37]}{C. Roman, Expanding the ${}^{171}\rm{Yb}^{+}$ toolbox: The ${}^{2}F^{\rm{o}}_{7/2}$ state as resource for quantum information science, Ph.D. thesis,University of California, 2021.}

\bibitem{Brage2009}
{T. Brage, M. Andersson, and R. Hutton, ICAMDATA-2008: Sixth International Conference on Molecular Data and Their Applications, AIP Conf. Proc. No. 1125 (AIP, Melville, NY,2009), p. 18.}


\bibitem{Andersson2009}
\href{https://doi.org/10.1088/0031-8949/2009/T134/014021} {M. Andersson, Phys. Scr. \textbf{T134}, 014021 (2009).}


\bibitem{Li2011}
\href{https://doi.org/10.1016/j.physleta.2010.12.081} { J. G. Li, C. Z. Dong, P. J$\rm{\ddot{o}}$nsson, and G. Gaigalas, Phys. Lett. A \textbf{375}, 914 (2011)}







\end{thebibliography}
\end{document}